\documentclass[twocolumn]{aastex631}
\usepackage{color}
\usepackage{amsmath}

\newcommand{\ie}{i.e.,\ }
\newcommand{\eg}{e.g.,\ }

\newcommand{\kms}{km~s$^{-1}$}
\newcommand{\Msun}{M$_\sun$}

\begin{document}

\title{A Dynamical Model of the M101/NGC~5474 Encounter}

\author{Sean T. Linden}
\affiliation{Department of Astronomy, University of Massachusetts at Amherst, Amherst, MA 01003, USA}

\author{J. Christopher Mihos}
\affiliation{Department of Astronomy, Case Western Reserve University, Cleveland OH 44106, USA}

\begin{abstract}

We present the first dynamical simulation that recreates the major
properties of the archetypal nearby spiral galaxy M101. Our model
describes a grazing but relatively close ($\sim$ 14 kpc) passage of the
companion galaxy NGC~5474 through M101's outer disk approximately 200
Myr ago. The passage is retrograde for both disks, which yields a
relatively strong gravitational response while suppressing the formation
of long tidal tails. The simulation reproduces M101's overall
lop-sidedness, as well as the extended NE Plume and sharp western edge
of the galaxy's disk. The post-starburst populations observed in M101's
NE Plume are likely a result of star formation triggered at the point of
contact where the galaxies collided. Over time, this material will mix
azimuthally, leaving behind diffuse, kinematically coherent stellar
streams in M101's outer disk. At late times after the encounter, the
density profile of M101's disk shows a broken ``upbending'' profile
similar to those seen in spiral galaxies in denser environments, further
demonstrating the connection between interactions and long-term
structural changes in galaxy disks.

\end{abstract}


\section{Introduction}

Due to its proximity and large angular size, the spiral galaxy M101 is
one of the most well-studied galaxies in the sky, yielding important
information on the dynamics of spiral structure, the interplay between
ISM physics and star formation, and the formation and evolution of stars
and stellar populations. However, M101's strong global asymmetry and
complex HI kinematics \citep[\eg][]{Waller97, Walter08, Mihos12,
Mihos13} show that the the galaxy is not in dynamical equilibrium,
complicating its use as a template for studying the physical processes
at work in galaxy disks.

While interactions in the group environment have clearly shaped M101's
recent history \citep[\eg][]{Waller97}, the details are poorly
constrained, and even the main interaction partner remains unclear. The
dwarf galaxy NGC~5477 lies near the edge of M101's disk, but is likely
too low in mass to drive M101's highly asymmetric HI structure.
Meanwhile, the brighter companion NGC~5474 is projected $\sim$ 90 kpc to
the southeast, and while the galaxy shows an off-center bulge, its disk
is otherwise normal kinematically and morphologically \citep{Rownd94}.
No detailed model exists describing M101's interaction history, and this
uncertainty makes it hard to link the dynamical conditions in the galaxy
to the properties of its ISM and exising stellar populations.

However, recent data have provided new constraints to our understanding
of the M101 system. Deep imaging of M101's outer disk has revealed
additional signatures of a recent interaction, including the extended NE
Plume and E Spur \citep{Mihos13}, but no evidence for long tidal tails
or connecting tidal bridges between M101 and any of its companions.
Meanwhile, {\it Hubble} imaging of the stellar populations in M101's NE
Plume has revealed $\sim$ 300 Myr old post-starburst population
\citep{Mihos18}, likely marking the time since M101's most recent
interaction. These observations, coupled with deep HI mapping of the
M101 system \citep{Walter08, Mihos12, Xu21} and new imaging and
spectroscopy of NGC~5474 \citep{Bellazini20} provide updated constraints
on a potential interaction between M101 and NGC~5474. We use these new
constraints to develop the first self-consistent N-body simulation of
the M101/NGC~5474 encounter, explaining a wide variety of the observed
properties while also giving insight into the subsequent evolution of
M101's disk. Given M101's important role as a template for studying the
detailed physical processes governing disk evolution, ISM physics, and
star formation in spiral galaxies, as well as its status as the dominant
spiral in a galaxy group \citep{Geller83, Tully88}, this simulation sets
the stage not only for a better understanding of M101 specifically, but
more generally for spiral galaxies and their evolution in group
environments.

\section{Simulation Methods}

\begin{figure*}[]
\centerline{\includegraphics[width=7.25truein]{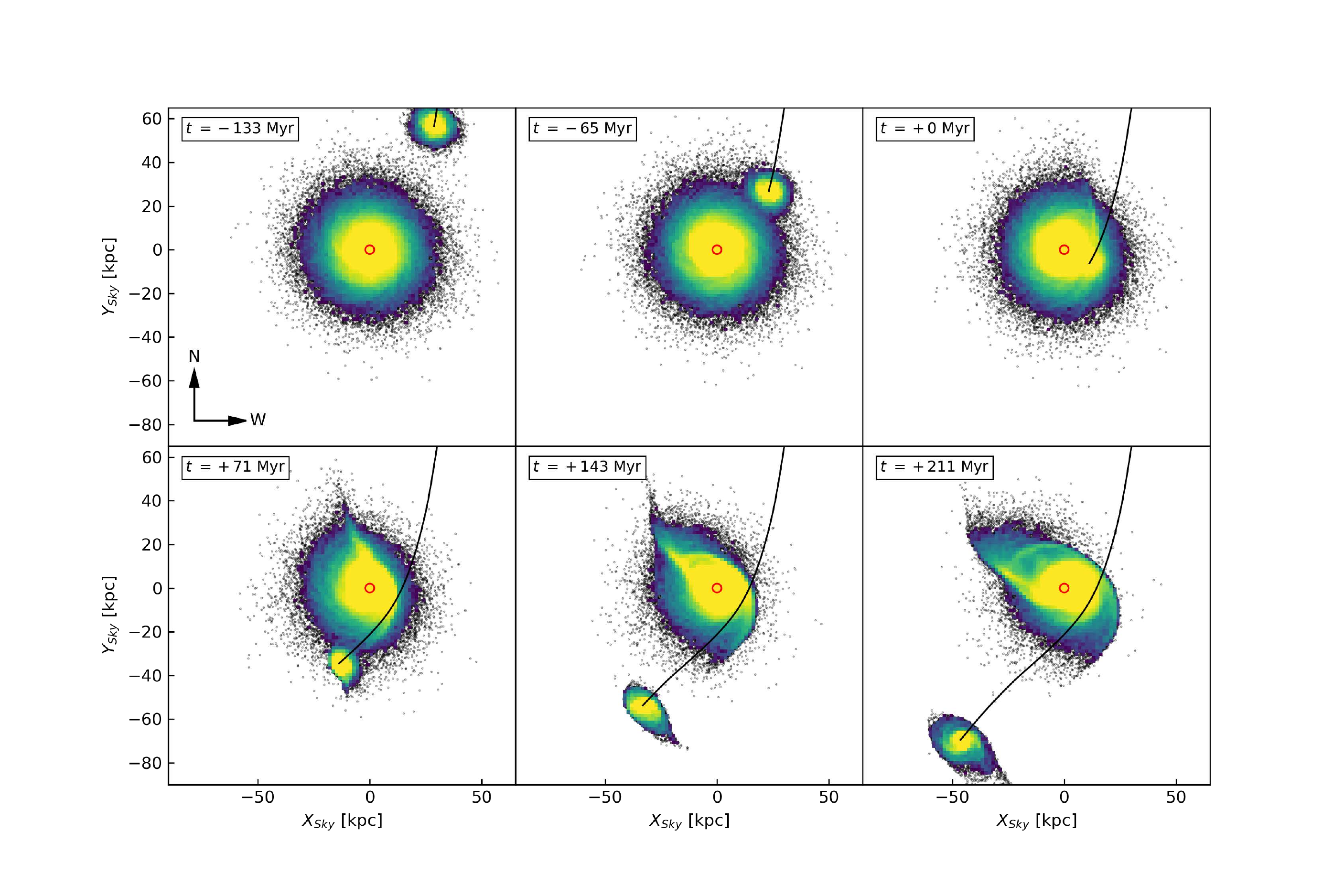}}
\caption{Sky view of the M101/NGC~5474 encounter. Time is measured
relative to the moment of periapse ($t=0$), with the last panel showing
the current time. The red circle shows a 2 kpc radius centered on M101's
nucleus. The simulation is visualized using a frame of reference that
fixes M101 at the origin at all times, and NGC~5474's orbital track is
shown by the black line.
An animation of this figure is available, showing the evolution of the interaction 
from $t=-269$ Myr to $t=+374$ Myr. The real-time duration of the animation is 
24 seconds, and the animation pauses for five seconds at $t=+211$ Myr,
the time of best match to the present-day M101 / NGC 5474 system.
}
\label{EvolSkyPlane}
\end{figure*}

While the focus of this study is the interaction between M101 and its
companion galaxy NGC~5474, in a preliminary set of models we also
examined the effects of M101's smaller companion, NGC~5477. We ran
simulations involving low mass companions ($M_{\rm N5477}/M_{\rm M101} =
0.01$, comparable to the B-band luminosity ratio of the galaxies) on a
variety of circular and elliptical orbits chosen to match the 44 kpc
projected separation of the pair. While these simulations showed a
propensity to drive symmetric two-armed spiral modes in M101, none were
successful in reproducing M101's strong $m=1$ disk asymmetry, or the
observed tidal morphology of its outer disk. These outcomes led us to
turn to simulations involving a stronger fly-by encounter between M101
and the more distant and more massive companion NGC~5474, described
below.

To build our N-body galaxy models of M101 and NGC~5474, we follow the method of
\citet{Hernquist93}, wherein each galaxy consists of a stellar disk, a
surrounding dark matter halo, and, in the case of the NGC~5474 model, a
central bulge. We omit the bulge component in the M101 model due to the
galaxy's extremely low bulge:disk ratio \citep[\eg][]{Kormendy10}. The
disks follow an exponential density profile, while both the dark halos
and the NGC~5474 bulge follow a spherical \citet{Hernquist90} model:
$$\rho(r) = {M \over {2\pi}} {a \over r} {1 \over (r+a)^3}.$$ We set the
halo scale radius ($a_h$) for the M101 model to be 10 times the disk
scale length, and, for computational expediency, truncate the dark halo
beyond 100 disk scale lengths. The disk:halo mass ratio of the M101
model is 1:23, and we set the circular velocity and disk scale length of
the M101 model to match M101's observed properties ($V_{c,\rm M101} = 220$
\kms\ and $h_{R,\rm M101}=4.4$ kpc; \citealt[][respectively]{Bosma81,
Mihos13}). This yields a total disk mass of
$M_{disk}=5.3\times10^{10}$~\Msun, and gives the model a slowly rising
rotation curve that flattens at large radius.

We build the NGC~5474 model in a similar fashion, setting its disk scale
length to $h_{R,\rm N5474}=1.5$ kpc and adding a central bulge with a
bulge:disk mass ratio of 1:3, in rough accordance with the observed
properties of the system \citep[\eg][]{Bellazini20}. However, in reality
NGC~5474's bulge is strangely offset from the center of its disk by
$\sim$ 1 kpc \citep[\eg][]{Rownd94,Pascale21}, suggesting the galaxy is
likely well out of equilibrium, and our attempts to match the observed
kinematics of the galaxy proved problematic. Scaling our galaxy model in
mass to match the NGC~5474's observed circular velocity \citep[40
\kms;][]{Rownd94}, gives a total (disk+bulge+halo) mass of
$2.5\times10^{10}$~\Msun\ and a mass ratio of the M101/NGC~5474 system
of $\approx$ 35:1. However, in preliminary models at this mass ratio, we
found that NGC~5474 to be too low in mass to drive the strong asymmetric
response seen in M101's disk. Instead, we were pushed to a higher mass
for the companion, ultimately adopting an 8:1 mass ratio for the galaxy
pair, which yielded a significantly higher circular velocity for
NGC~5474 of 135 \kms. This mass discrepancy remains one shortcoming of
our model, but we note that recent simulations of NGC~5474 by
\citet{Pascale21} that try to match both the morphology and kinematics
of this galaxy suggest that the NGC~5474's bulge and disk may in fact be
two separate and interacting galaxies themselves. Given the surprisingly
complex nature of this galaxy, and its uncertain dynamical state, we
choose to focus our modeling efforts largely on M101's response, and
leave the NGC~5474 discrepancy for future follow up studies.

Our primary goal in designing the encounter model was to reproduce the
large scale morphology and kinematics of M101: the galaxy's strong $m=1$
lopsidedness, its extended NE Plume, and the sharp edge to the western
side of its disk \citep{Mihos13}. These are all gravitational effects,
so for simplicity our simulations are purely collisionless and do not
model the hydrodynamic evolution of the interstellar medium or the
star-forming response of the disk. We evolve the simulations using the
hierarchical treecode of \citet{Hernquist87}, with fixed timestep of
0.68 Myr and using a total of 1,00,000 and 165,000 particles to model
M101 and NGC~5474, respectively.

Aside from the morphological constraints, our simulations are also
constrained by the projected relative positions and velocities
of the galaxies. These various considerations led to limitations on
orbital geometries of the disks \citep[disk inclinations $i$ and
arguments of periapse $\omega$, see][]{TT72} and viewing angle of the
system. The small difference in systemic velocity (30 \kms) led us to
focus on orbits that occurred largely in sky plane, to minimize the
projected orbital velocity. Because M101 and NGC~5474 are both observed
mostly face-on, this argument in turn implied highly prograde or
retrograde orbits ($i \sim 0\degr$ or $180\degr$), rather than polar
encounters ($i\sim90\degr$). In our initial modeling tests we found that
prograde encounters strong enough to drive the observed M101 asymmetry
also resulted in long tidal tails, which are not observed in the system.
These tests also showed that the resulting lopsidedness of M101's disk
was a strong function of pericenter distance, while M101's argument of
periapse $\omega$ was limited to a small range due to the need to match
the current projected location of NGC~5474 relative to M101. All these
considerations ultimately led us to favor the dual retrograde encounter
described below.

\begin{figure*}[]
\centerline{\includegraphics[width=7.25truein]{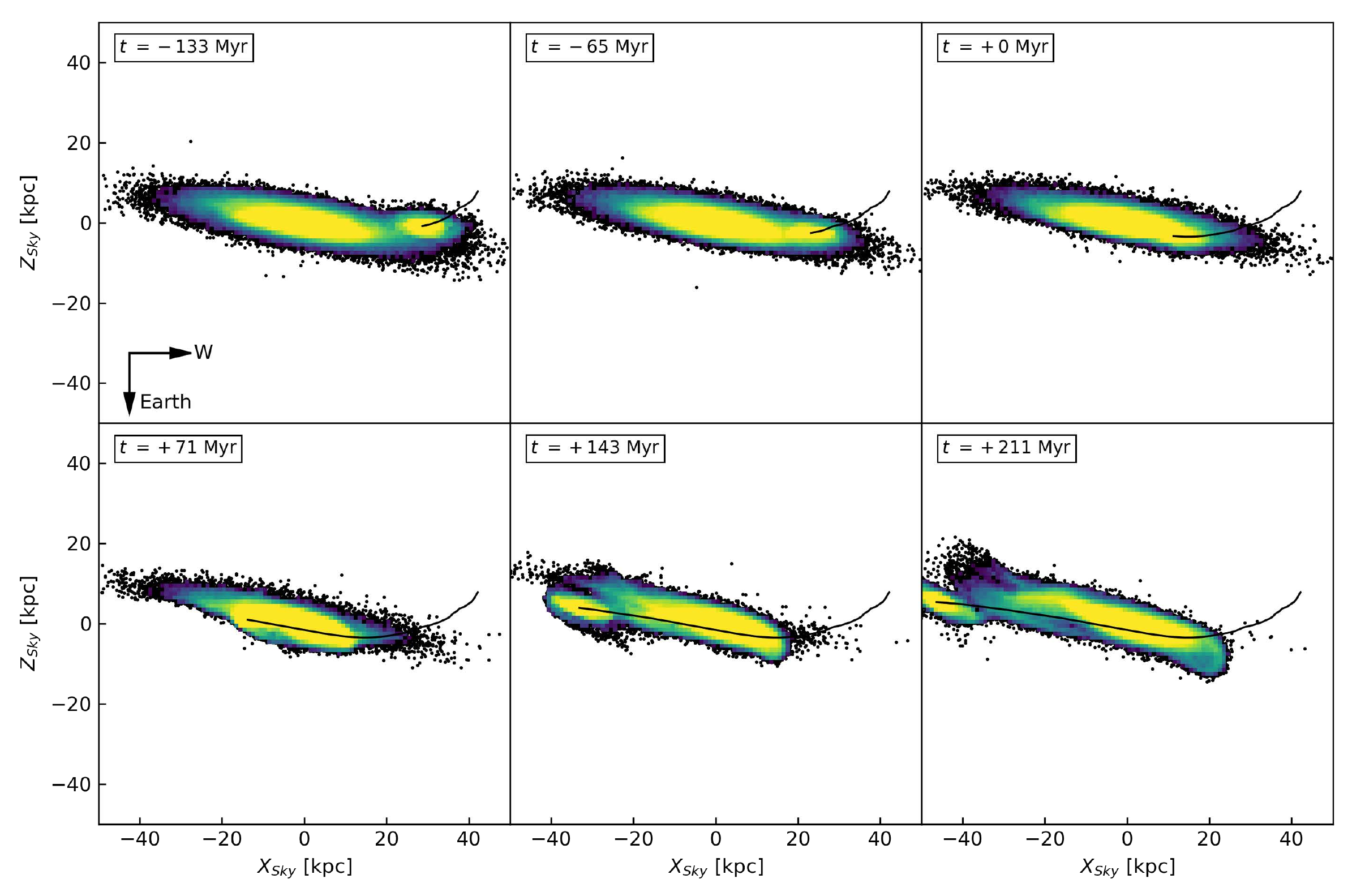}}
\caption{The M101/NGC~5474 simulation viewed orthogonal to the sky
plane. Time is measured relative to the moment of periapse ($t=0$), with
the last panel showing the current time. The simulation is visualized
using a frame of reference that fixes M101 at the origin at all times,
and NGC~5474's orbital track is shown by the black line.
An animation of this figure is available, showing the evolution of the interaction 
from $t=-269$ Myr to $t=+374$ Myr. The real-time duration of the animation is 
24 seconds, and the animation pauses for five seconds at $t=+211$ Myr,
the time of best match to the present-day M101 / NGC 5474 system.
}
\label{EvolOrthoPlane}
\end{figure*}

\section{Best-match Interaction Scenario}

Our best-match simulation begins with NGC~5474 located 175 kpc away from
M101, moving on a parabolic orbit initialized as Keplerian with a 4.4
kpc pericenter distance. However, due to the galaxies' extended halo
mass distributions, the orbital track deviates from this idealized
orbit, and the galaxies reach a much wider periapse of 13.7 kpc. The
orientation of both disks is primarily retrograde, with inclinations and
arguments of periapse given by $(i,\omega)_{\rm M101}=(195\degr,265\degr)$
and $(i,\omega)_{\rm N5474}=(150\degr,100\degr)$. The system is viewed with
the orbit plane slightly inclined by $\sim 10$\degr\ from the sky plane
($\sim 15$\degr\ from M101's disk plane).\footnote{Additional simulation
details, particle snapshot files, and visualizations are available at
\href{http://astroweb.case.edu/hos/M101Sim}{http://astroweb.case.edu/hos/M101Sim}.
}

As the interaction proceeds (Figure~\ref{EvolSkyPlane}), NGC~5474 enters from the
north, passing through the west side of M101's outer disk
approximately 200 Myr ago. Material on the side of the disk near the
impact point feels a dispersive impulse and is scattered outwards,
eventually rotating around the galaxy to be seen today as the NE Plume.
Meanwhile, the passage of the companion pulls M101 westward, leading to
crowding and compression of material along the disk's current western
edge. The interaction is not so strong as to lead to a merger; NGC~5474
exits to the southeast and continues to move away from M101 today.
Viewed edge-on to the sky plane (Figure~\ref{EvolOrthoPlane}), we see the oblique
nature of the encounter. NGC~5474 skims through M101's outer disk at a
shallow 15\degr\ angle, scattering material near the contact point vertically out
of the disk plane. This leads to a flaring of M101's outer disk, with material
currently in the NE Plume moving away along our line of sight (\ie
``behind'' M101's disk plane from our perspective).

\begin{figure*}[]
\centerline{\includegraphics[width=7.25truein]{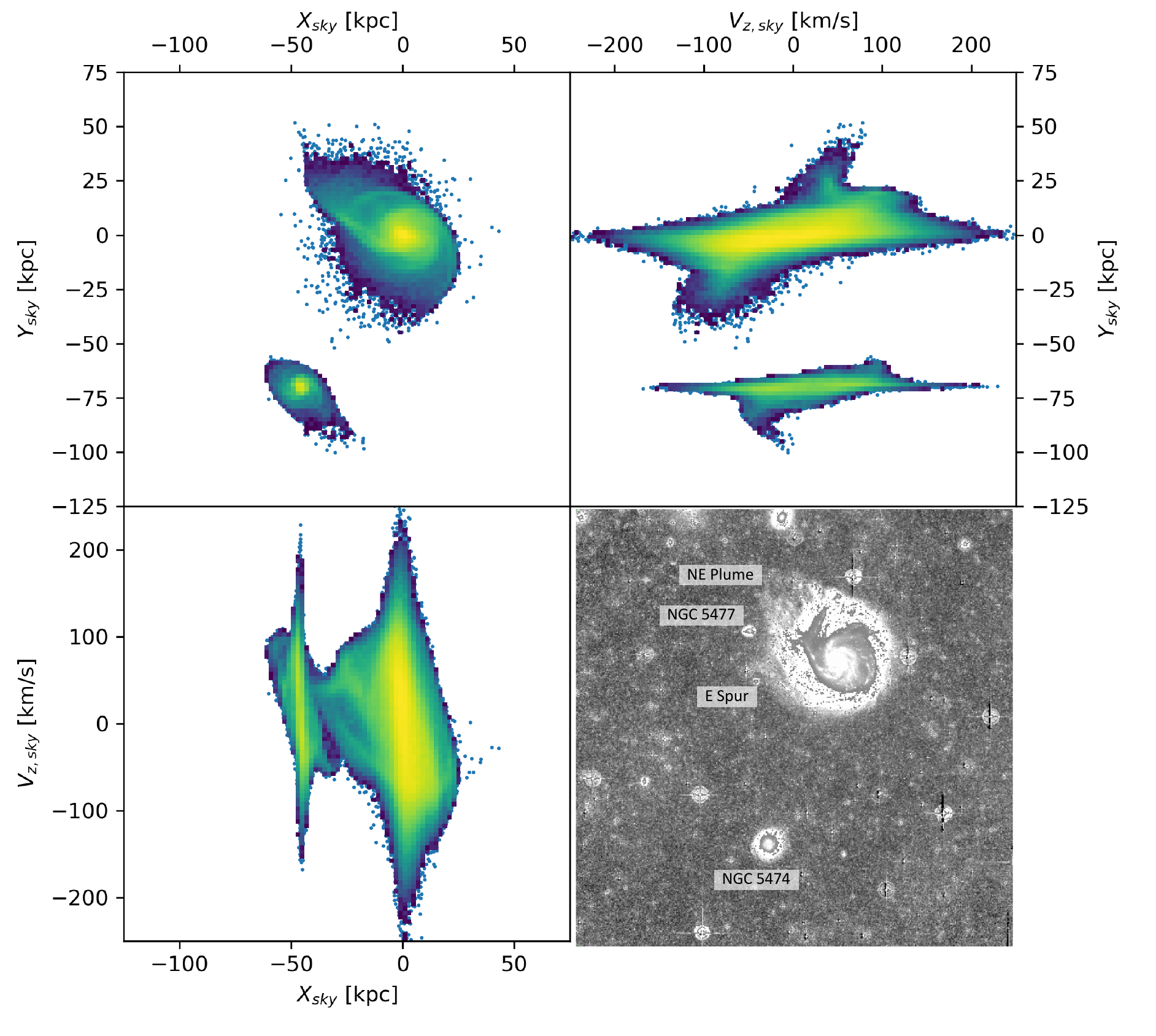}}
\caption{Position-velocity plots of the M101-NGC~5474 interaction model
observed at the present time. The top left panel show the sky view, while the
top right and lower left panels show the model velocities collapsed along the
X and Y coordinates, respectively. Radial velocity is measured relative to
M101's systemic velocity. The lower right panel shows a deep B-band image
of M101 from \citet{Mihos13}, with companion galaxies and tidal features marked.}
\label{PVimage}
\end{figure*}

The simulation reproduces much of the morphology of the M101 system,
including the lopsidedness of the inner disk, the extended NE Plume, and
the sharp isophotal cutoff on the disk's western edge. M101's prominent
NE arm can be seen developing shortly after the initial impact, then
swinging around to its current location; we also see a broken dog-leg
structure that bends the arm to the northwest, similar to that observed
in M101 itself (see Figure~\ref{PVimage}). The relative position and velocity of
NGC~5474 is reproduced well, while the dual retrograde geometry explains
the lack of long tidal tails around either galaxy.

\begin{figure*}[]
\centerline{\includegraphics[width=7.25truein]{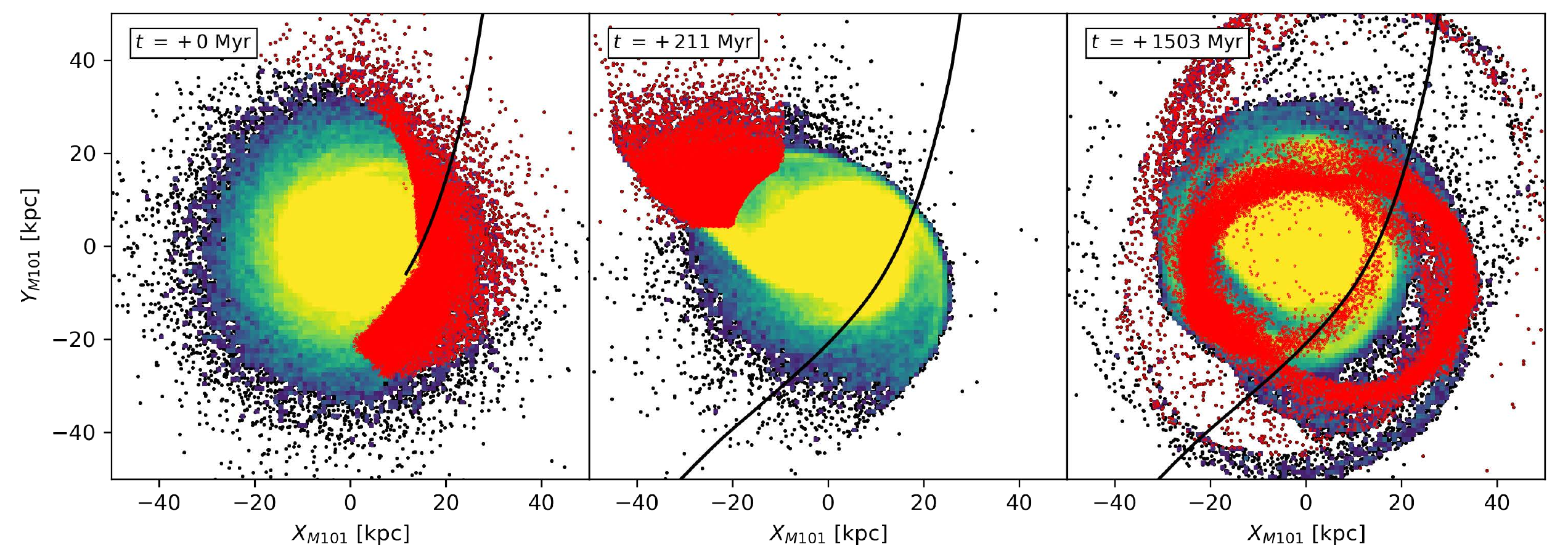}}
\caption{Zoomed in view of the M101 disk at the moment of periapse
(left), the current time (center), and 1.3 Gyr in the future (right). NGC
5474's orbital track is shown in black, and in each panel, particles
found in today's NE Plume are shown in red. At periapse, material in the
NE Plume was on the west side of M101, near the contact point of
NGC~5474, but has since rotated around to form today's NE Plume. In
another Gyr, this material will have mixed azimuthally but remain in a
coherent stream located in M101's outer disk.
An animation of this figure is available. The animation shows the evolution
of M101's disk from $t=-469$ Myr to $t=+2064$ Myr, with particles
in today's NE Plume shown in red. The real-time 
duration of the animation is 15 seconds.
}
\label{PlumeEvol}
\end{figure*}

Of particular interest is the history of NE Plume. Its very blue
integrated colors led \citet{Mihos13} to propose a recent weak burst of
star formation in the region. This was recently confirmed by deep {\sl
Hubble Space Telescope} imaging \citep{Mihos18}, which revealed the
presence of $\sim$ 300 Myr old post-starburst stellar populations in the
Plume. According to our simulation, material in today's NE Plume was
originally at the contact point where NGC~5474 passed through M101's
disk, as shown in Figure~\ref{PlumeEvol}. As NGC 5474 approaches M101,
compressive tidal forces should lead to an increase in the turbulent
energy and the velocity dispersion of the ISM. Hydrodynamical
simulations of starburst activity in galaxy mergers suggest that the
kinetic energy carried by compressive turbulence, and the subsequent
increase in the dense gas fraction in the ISM begins as early as $\sim
50$ Myr before closest approach \citep{Renaud14,Renaud18}. Indeed,
in our own simulations shown here, an analysis of the local velocity
field in M101's outer disk near the collision point shows signatures of
convergent velocity 25--50 Myr before the moment of periapse, as the
companion comes in contact with M101's disk \citep[see
also][]{Renaud09}. While our collisionless simulation precludes us from
following the hydrodynamical response of the ISM, the enhanced
turbulence driven by these compressive effects likely generates an
excess of dense gas at this location, triggering the starburst episode
recorded in the NE Plume's stellar populations. Therefore, depending on
the early response of the star-forming gas, we might expect to see
differences of 50 Myr between the $\sim$ 300 Myr old stellar population
age in the Plume and the $\sim$ 200 Myr time since periapse in our
simulation, which partially explains these slightly discrepant
timescales. These details aside, the dynamical connection between the
Plume's post-starburst stellar population and star formation triggered
at the original impact site is plausibly demonstrated in our model.

Deep imaging of M101's outer disk has also revealed a second tidal
feature, the fainter and redder ``Eastern Spur'' \citep{Mihos13},
extending from the SE side of M101's disk. While our simulation
shows no distinct feature here, material on this side of the disk was
originally on the north side  of M101 and moving away from NGC~5474
when the companion passed through M101's disk. Thus, this material 
felt a much weaker perturbation, likely explaining its redder colors, 
lack of induced star formation, and more passive evolution.

Our simulation also captures the observed kinematics of the
M101/NGC~5474 system quite well. Figure~\ref{PVimage} shows position-velocity
plots of the simulation viewed at the best-match time, where it can be
seen that NGC~5474 has a small +20 \kms\ redshift relative to M101,
similar to the observed value of $32\pm11$ \kms\ \citep{Zaritsky90,
Rownd94}. The model also has the sense of rotation in M101's disk
correct, with the north side being redshifted relative to systemic
\citep{Walter08}. More detailed kinematic matching to existing HI
velocity data proves difficult, given that our simulation lacks any
hydrodynamic component. Nonetheless, some broad inferences are possible.
Deep 21-cm mapping of the system \citep{Huchtmeier79, Mihos12, Xu21} has
revealed HI between the two galaxies at intermediate velocities. Our
model shows no similar {\it stellar} tidal bridge linking the galaxies,
which suggests the observed HI stream may be a result of ram-pressure
effects rather than simple tidal stripping. Similar arguments can be
made regarding the high-velocity gas associated with the NE Plume
\citep{vdh88, Walter08}, with velocities redshifted by $\gtrsim$ 100
\kms\ compared to M101's disk. We do see a spray of (collisionless)
material in the NE Plume extending to higher velocity (Figure~\ref{PVimage})
which might plausibly be associated with that high velocity cloud
structure. This is material scattered out of the disk by NGC~5474's
passage, and ram-pressure stripping might accentuate the structures seen
here. However, at the moment of periapse, the line-of-sight velocity of
the companion as it impacted M101's disk was only $\sim$ 60 \kms,
significantly lower than the 100 \kms\ velocity spread of the high
velocity gas now seen in the NE Plume. It remains unclear, therefore,
whether ram-pressure effects could explain the full velocity range of
the gas. Follow-up simulations that include hydrodynamics will be useful
to test this aspect of the scenario.

\section{Future Evolution}

With M101's encounter with NGC~5474 only $\sim$ 200 Myr old --- less
than a single rotation period for M101's outer disk --- and its disk
showing such strong, global asymmetry, the galaxy's future evolution is
very much an open question. Here we evolve the simulation forward in
time to evaluate the effect of the encounter on M101's long-term
evolution. The encounter has led to only a modest amount of orbital
decay in the system, leaving NGC~5474 on a very wide and loosely bound
orbit that will not produce another close passage for many Gyr. Thus
over long time scales, M101's further evolution is largely governed by
its return to equilibrium after the initial perturbation of NGC~5474's
passage.

Figure~\ref{PlumeEvol}c shows the M101 model approximately 1.3 Gyr after the
impact. At this point, material in today's NW Plume has wrapped
completely around the galaxy, forming a coherent stream spanning a
radial extent of $\approx$ 15--40 kpc. Given that the stellar population
of the Plume is largely coeval \citep{Mihos18}, the stream will not only
be kinematically coherent, but also distinct in age and metallicity from
the bulk of M101's stellar populations. Over time, this stream will
continue to slowly mix radially and azimuthally in M101's outer disk,
but --- barring scattering or disruption by subsequent interactions
(such as with the dwarf satellite NGC~5477) --- will remain distinct in
terms of its kinematics and stellar populations. Such a long-lived
coherent stellar stream might be hard to distinguish from true accretion
streams formed by the disruption of infalling satellite galaxies. Given
the ubiquitous presence of star streams in the Milky Way and other
spiral galaxies, it is possible that some of these observed streams may
actually be due to original disk material ``returning home'' after a
tidal encounter with a neighboring companion.

More globally, we can also examine changes in the radial distribution of
stars in M101's disk. Disk galaxies often show differences in the radial
scale length between the inner and outer disk, including both upbending
($h_{R,\rm outer} > h_{R,\rm inner}$) and downbending ($h_{R,\rm outer} <
h_{R,\rm inner}$) radial profiles \citep[\eg][]{Erwin05, Pohlen06}. Growing
evidence links upbending profiles to galaxies in denser environments
\citep{Pohlen06, Watkins19}, suggesting interactions can either scatter
disk material outwards, or deposit accreted stars into a galaxy's outer
disk. \citet{Mihos12} showed that M101's surface brightness profile
shows complex azimuthal variations (both downbending and upbending
profiles) outside of a radius of $\approx$ 8\arcmin\ (16 kpc). That
radius is similar to the pericenter distance of our encounter model (14
kpc), suggesting that the interaction is indeed reshaping M101's outer
disk. If we measure the M101 disk profile at late times ($t\approx$
1.5--2 Gyr), after the present-day asymmetries have had time to mix
azimuthally, we find evidence for spreading of the outer disk. Prior to
the encounter encounter, the disk had a uniform exponential profile with
scale length $h_{R}=4.4$ kpc, but at late times after the encounter, the
disk shows an upbending profile with inner ($R<15$ kpc) and outer
($R>15$ kpc) scale lengths of $h_{R}=4.2\pm0.1$ kpc and $5.3\pm0.2$ kpc,
respectively. Thus, this interaction-driven reshaping of disk profiles
may be a common evolutionary outcome in the loose group environment.

\section{Summary and Future Directions}

We have constructed the first N-body simulation to capture the
gravitational encounter between M101 and its companion NGC~5474. The
encounter involves a grazing, low inclination passage of NGC~5474
through M101's outer disk ($R_{peri} \approx 14$ kpc), with closest
approach occurring approximately 200 Myr ago. The interaction was a
retrograde passage for both galaxies, taking place largely in the plane
of the sky as observed from our vantage point. Our simulation reproduces
many of the features observed in the M101 system today: including the
projected distance and velocity of the two galaxies, M101's overall
lopsided asymmetry, the morphology of M101's NE spiral arm and NE Plume,
and the sharp edge to the west side of M101's disk. We also roughly
match the 200 -- 300 Myr timescale for the encounter inferred from
stellar age constraints in M101's NE Plume, and show that this stellar
population was likely formed in a burst of star formation triggered at
the contact point when the two disks originally collided.

Evolving the system forward in time to evaluate the long-term evolution
of M101, we find that no merger is imminent; NGC~5474 continues to move
away from M101 on a wide and very loosely bound orbit, while the
asymmetric structures currently seen in M101's disk slowly mix
azimuthally as the galaxy recovers from the collision. Material
currently in M101's NE Plume falls back wraps around the galaxy,
resulting in a discrete stellar stream that is both phase-space coherent
and coevol in age. At late times, M101's disk also shows an
``upbending'' surface density profile due to inner disk material drawn
out by the interaction, similar to the upbending profiles often seen in
disk galaxies found in dense group and cluster environments.

While our simulation reproduces many features of the M101 system, there
remains room for improvement. First, our use of a collisionless
simulation reproduces only the large scale gravitational response to the
encounter; adding gas physics to the simulation would let us examine the
hydrodynamical response, including the detailed structure of the M101's
spiral arms, the star forming response of the disk, and the origin of
M101's high velocity gas. Second, further adjustments to the interaction
model -- such as a slower encounter velocity and a lower companion mass
-- might fix both our need for an overly-massive model for NGC~5474 and
the slight ($\sim$ 100 Myr) mismatch between the simulated collision
time and the age of the post-starburst population in M101's outer disk.
And finally, our simulation also ignores any role played by the dwarf
galaxy NGC~5477. While this system is almost certainly too low mass to
drive the large scale asymmetry of M101, it could affect the outer disk
morphology that constrains our model. Nonetheless, given the overall
uncertainties, we find our interaction model to be a promising new
description of M101's recent interaction history and its subsequent
evolution.

\begin{acknowledgements}

The authors thank George Privon, Stacy McGaugh, and Ray Garner for help 
and encouragement over the course of this study.

\end{acknowledgements}

\bibliographystyle{aasjournal}

\begin{thebibliography}{}

\bibitem[Bellazzini et al.(2020)]{Bellazini20} Bellazzini, M., Annibali, F., Tosi, M., et al.\ 2020, \aap, 634, A124. 
\bibitem[Bosma et al.(1981)]{Bosma81} Bosma, A., et al.\ 1981, \aap, 93, 106.
\bibitem[Erwin et al.(2005)]{Erwin05} Erwin, P., Beckman, J.~E., \& Pohlen, M.\ 2005, \apjl, 626, L81. 
\bibitem[Geller \& Huchra(1983)]{Geller83} Geller, M.~J. \& Huchra, J.~P.\ 1983, \apjs, 52, 61.
\bibitem[Kormendy et al.(2010)]{Kormendy10} Kormendy, J.,  et al.\ 2010, \apj, 723, 54.
\bibitem[Mihos et al.(2013)]{Mihos13} Mihos, J.~C.,  et al.\ 2013, \apj, 762, 82. 
\bibitem[Mihos et al.(2012)]{Mihos12} Mihos, J.~C., et al.\ 2012, \apj, 761, 186. 
\bibitem[Mihos et al.(2018)]{Mihos18} Mihos, J.~C., et al.\ 2018, \apj, 862, 99. 
\bibitem[Hernquist(1987)]{Hernquist87} Hernquist, L.\ 1987, \apjs, 64, 715.
\bibitem[Hernquist(1993)]{Hernquist93} Hernquist, L.\ 1993, \apjs, 86, 389. 
\bibitem[Hernquist(1990)]{Hernquist90} Hernquist, L.\ 1990, \apj, 356, 359. 
\bibitem[Huchtmeier \& Witzel(1979)]{Huchtmeier79} Huchtmeier, W.~K. \& Witzel, A.\ 1979, \aap, 74, 138.
\bibitem[Pohlen \& Trujillo(2006)]{Pohlen06} Pohlen, M. \& Trujillo, I.\ 2006, \aap, 454, 759. 
\bibitem[Pascale et al.(2021)]{Pascale21} Pascale, R., Bellazzini, M., Tosi, M., et al.\ 2021, \mnras, 501, 2091.
\bibitem[Renaud et al.(2009)]{Renaud09} Renaud, F., Boily, C.~M., Naab, T., et al.\ 2009, \apj, 706, 67. 
\bibitem[Renaud et al.(2014)]{Renaud14} Renaud, F., Bournaud, F., Kraljic, K., et al.\ 2014, \mnras, 442, 33.
\bibitem[Renaud et al.(2018)]{Renaud18} Renaud, F., Athanassoula, E., Amram, P., et al.\ 2018, \mnras, 473, 585.
\bibitem[Rownd et al.(1994)]{Rownd94} Rownd, B.~K., et al.\ 1994, \aj, 108, 1638.
\bibitem[Toomre \& Toomre(1972)]{TT72} Toomre, A. \& Toomre, J.\ 1972, \apj, 178, 623.
\bibitem[Tully(1988)]{Tully88} Tully, R.~B.\ 1988, Cambridge: University Press, 1988
\bibitem[van der Hulst \& Sancisi(1988)]{vdh88} van der Hulst, T. \& Sancisi, R.\ 1988, \aj, 95, 1354. 
\bibitem[Waller et al.(1997)]{Waller97} Waller, W.~H.,  et al.\ 1997, \apj, 481, 169. 
\bibitem[Walter et al.(2008)]{Walter08} Walter, F., Brinks, E., de Blok, W.~J.~G., et al.\ 2008, \aj, 136, 2563. 
\bibitem[Watkins et al.(2019)]{Watkins19} Watkins, A.~E., Laine, J., Comer{\'o}n, S., et al.\ 2019, \aap, 625, A36.
\bibitem[Xu et al.(2021)]{Xu21} Xu, J.-L., Zhang, C.-P., Yu, N., et al.\ 2021, \apj, 922, 53.
\bibitem[Zaritsky et al.(1990)]{Zaritsky90} Zaritsky, D., Elston, R., \& Hill, J.~M.\ 1990, \aj, 99, 1108.

\end{thebibliography}

\end{document}